\documentclass[%
 reprint,
amsmath,amssymb,
aps,
prx
]{revtex4-2}

\usepackage{graphicx}
\usepackage{dcolumn}
\usepackage{bm}
\usepackage{amsmath}
\usepackage{amssymb}

\newcommand{\cmt}[1]{{}}
\usepackage{xcolor}
\usepackage{times}

\usepackage{xcolor}

\begin{document}
\preprint{APS/123-QED}

\title{Broad Spectrum Coherent Frequency Conversion with Kinetic Inductance Superconducting Metastructures}


\author{Yufeng Wu}
\author{Chaofan Wang, Danqing Wang, Mingrui Xu, Yiyu Zhou}
\author{Hong X. Tang}%
 \email{hong.tang@yale.edu}
\affiliation{%
 Department of Electrical Engineering, Yale University, New Haven, Connecticut 06520, USA
}%

\date{\today}

\begin{abstract}
Parametric frequency converters (PFCs) play a critical role in bridging the frequency gap between quantum information carriers. PFCs in the microwave band are particularly important for superconducting quantum processors, but their operating bandwidth is often strongly limited. Here, we present a multimode kinetic metastructure for parametric frequency conversion between broadly spanning frequency modes. This device comprises a chain of asymmetric kinetic inductance grids designed to deliver efficient three-wave mixing nonlinearity. We demonstrate high efficient coherent conversion among broadly distributed modes, and the mode frequency is continuously tunable by controlling the external magnetic field strength, making it ideally suited for quantum computing and communication applications requiring flexible and efficient frequency conversion.
\end{abstract}

\maketitle

\section{Introduction}
Parametric frequency converters (PFCs) are important components for the interconnection of qubits, quantum memory and quantum sensors operating at distinct frequencies \cite{kimble2008quantum, simon2017towards, wehner2018quantum, awschalom2021development}. In particular, PFCs operating in the microwave frequency band are highly desired due to the wide application of microwave photons across various platforms, including superconducting qubits, rare-earth spins, mechanical resonators, and ferromagnetic resonators \cite{tabuchi2015coherent, chu2017quantum, moores2018cavity, landig2018coherent}. However, quantum microwave photons cannot propagate over long distances at room temperature due to the high losses and thermal noise in transmission lines. As a result, significant efforts have been dedicated to developing microwave-to-optical (M2O) converters, which link microwave signals to the optical telecom light for efficient long-distance transmission \cite{andrews2014bidirectional, rueda2016efficient, fan2018superconducting, forsch2020microwave, mckenna2020cryogenic, mirhosseini2020superconducting, fu2021cavity, xu2021bidirectional, han2021microwave}. For maximum M2O conversion efficiency, it is crucial, yet challenging, to align the microwave signal frequency with the microwave resonance frequency of an M2O converter. To achieve precise frequency alignment, quantum microwave-to-microwave (M2M) converters emerge as a promising solution. M2M converters have been previously implemented with Josephson parametric converters (JPC) \cite{zakka2011quantum, nguyen2012quantum, abdo2013full}, which exhibits near-unity efficiency and low added noise. However, these devices rely on the coupling between two spatial microwave modes and thus present limited conversion bandwidth and spectral coverage. By contrast, traveling-wave parametric amplifiers (TWPAs) can provide broadband gain \cite{esposito2022observation, perelshtein2022broadband} but face challenges in frequency conversion due to the mode selection issue. Specifically, a single pump tone can drive multiple conversion processes, which inherently prevents unity conversion efficiency \cite{peng2022floquet, malnou2024traveling}.

\begin{figure*}[t]
    \centering
    \includegraphics[width=\textwidth]{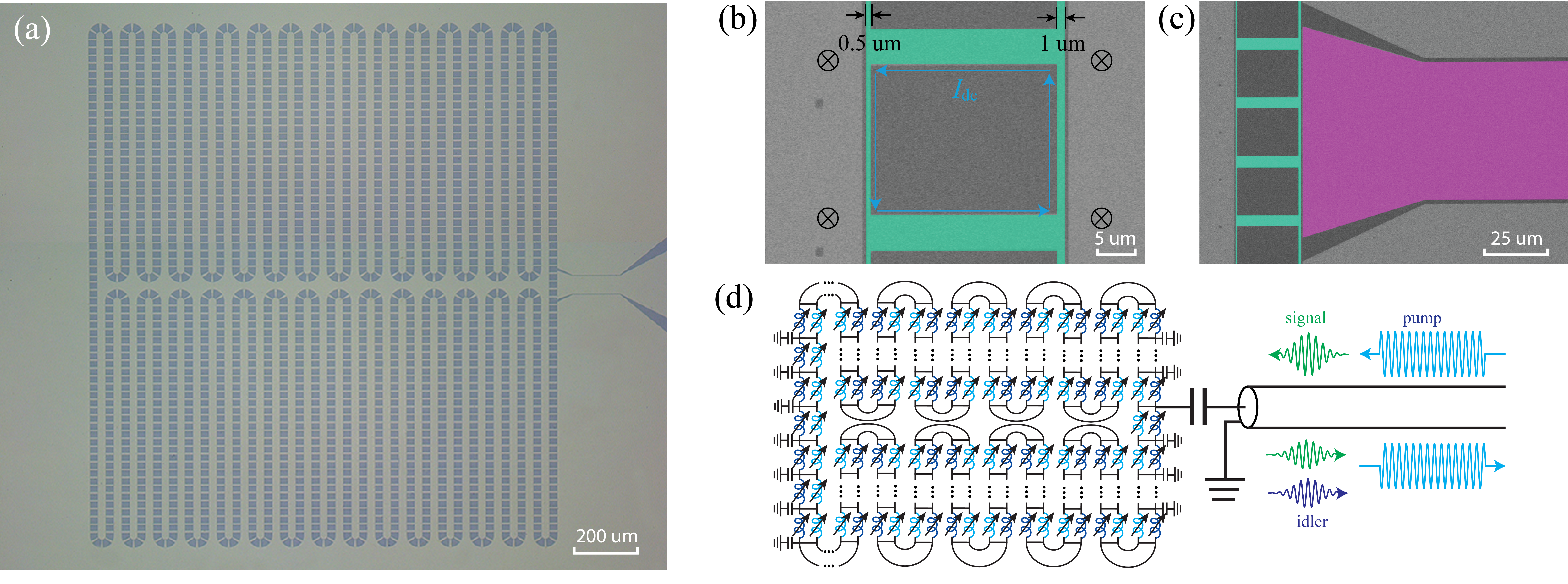}
    \caption{(a) Optical microscope image of the resonator, designed as a closed chain of microloops forming a compound ring resonator. The grey area represents NbN, while the darker area is the silicon substrate. (b) False-colored scanning electron microscope (SEM) image of a microloop, showing nanowires of asymmetric width. The arrows indicate the direction of the supercurrent induced by a remotely applied magnetic field. (c) False-colored SEM image of the coupling port of the resonator. The transmission line is colored pink. (d) Circuit diagram of the resonator. The interaction between the signal, idler, and pump tones within the resonator is illustrated by the differently colored waveforms.}
    \label{fig:1-device}
\end{figure*}

Here, we present a junction-free design that enables parametric frequency conversion across a broad spectrum of microwave frequencies. Our design features a kinetic inductance meta-ring formed by a ladder of microloops. The meta-ring is patterned on NbN thin film on silicon and supports whispering gallery modes. The left and right sides of each microloop comprise of a pair of nanowires of different widths, serving as kinetic inductance elements. In the presence of an out-of-plane magnetic field bias, this asymmetric microloop allows for the three-wave mixing Hamiltonian that is necessary for the conversion process. The magnetic field bias also enables frequency tuning \cite{xu2019frequency}. If the magnetic field induced frequency shift can exceed the free spectral range (FSR) of the ring, the device can match frequencies with arbitrarily selected microwave mode. In this work, we demonstrate 100\% FSR frequency tunability in the designed frequency band ($\sim$9.40~GHz), and 49\% tunability at an octave away ($\sim$ 4.85~GHz). We achieve ultra-high conversion efficiency with low thermal occupation in both signal and idler modes. The high conversion efficiency is further validated by the coherent interference between the converted idler tone and the reflected idler tone with up to 99.99\% interference visibility. The high conversion efficiency is maintained across different modes and varying magnetic field strengths. This platform thus holds great promise as a versatile quantum interface for hybrid quantum systems. 

\section{Device and Characterization}
Multimode resonators (MMRs) have been widely utilized in quantum information processing \cite{grezes2014multimode, naik2017random, sletten2019resolving, chakram2022multimode, matanin2023toward}. In contrast to single-mode resonators, the presence of a large number of modes in MMRs allows for phase matching over a broad band\cite{lei2023characterization}. MMRs have been experimentally demonstrated in various geometries, such as 3D cavities \cite{brecht2015demonstration, lei2023characterization}, half-wave and quarter-wave resonators \cite{pozar2011microwave}, and ring resonators \cite{mohamed2024selective}. Here, we use a multimode ring resonator for M2M conversion, which utilizes a ladder of microloops, as depicted in Fig.~\ref{fig:1-device} (a). The ladder is folded to provide a compact footprint. This design facilitates uniform external coupling rate among different modes, as opposed to quarter- or half-wave resonators. A micrograph image of a ladder section, shown in Fig. \ref{fig:1-device} (b-c), reveals a periodic microloop inspired by the Ouroboros frequency-tunable resonator \cite{xu2019frequency}. When exposed to an external magnetic field, a dc supercurrent $I_\mathrm{dc}$ is induced within the microloop, leading to a nonlinear modification of the inductance according to $L_k(I_\mathrm{dc}) = L_0\left[1 + (I_\mathrm{dc}/{I^*})^2\right]$, where $I^*$ represents the characteristic current of the nanobridge and is typically on the order of the critical current \cite{annunziata2010tunable}. Consequently, the resonance frequency is tuned by changes in kinetic inductance. To enhance the three-wave mixing Hamiltonian necessary for frequency conversion, the two nanowires along the rails of the ladder in the microloop are designed with different widths (see Appendix \ref{sec:app_freq_tuning} for detailed analysis), with the thinner wire being half the width of the wider one, as shown in Fig. \ref{fig:1-device} (b). These nanowires are the primary sources of kinetic inductance, as the bars of the ladder are much wider and contribute less kinetic inductance. The input signal and the pump tone are fed from and reflected back to a transmission line that is capacitively coupled to the device as shown in Fig. \ref{fig:1-device} (c). Figure \ref{fig:1-device} (d) illustrates the circuit diagram of the complete kinetic inductance metastructure. The nanowires and the nearby ground form distributed capacitors. The NbN plane inside the ring is galvanically isolated from the outer ground, and the flux-induced supercurrent in the inner plane is negligible compared to that generated from the microloop due to the smaller area-to-length ratio. The resonator is fabricated on a high-resistivity silicon substrate, with a 5 nm NbN film deposited via atomic layer deposition. Device patterns are defined using a single e-beam lithography step with CSAR resist, followed by fluorine dry etching. 

\begin{figure*}[t]
    \centering
    \includegraphics[width=\textwidth]{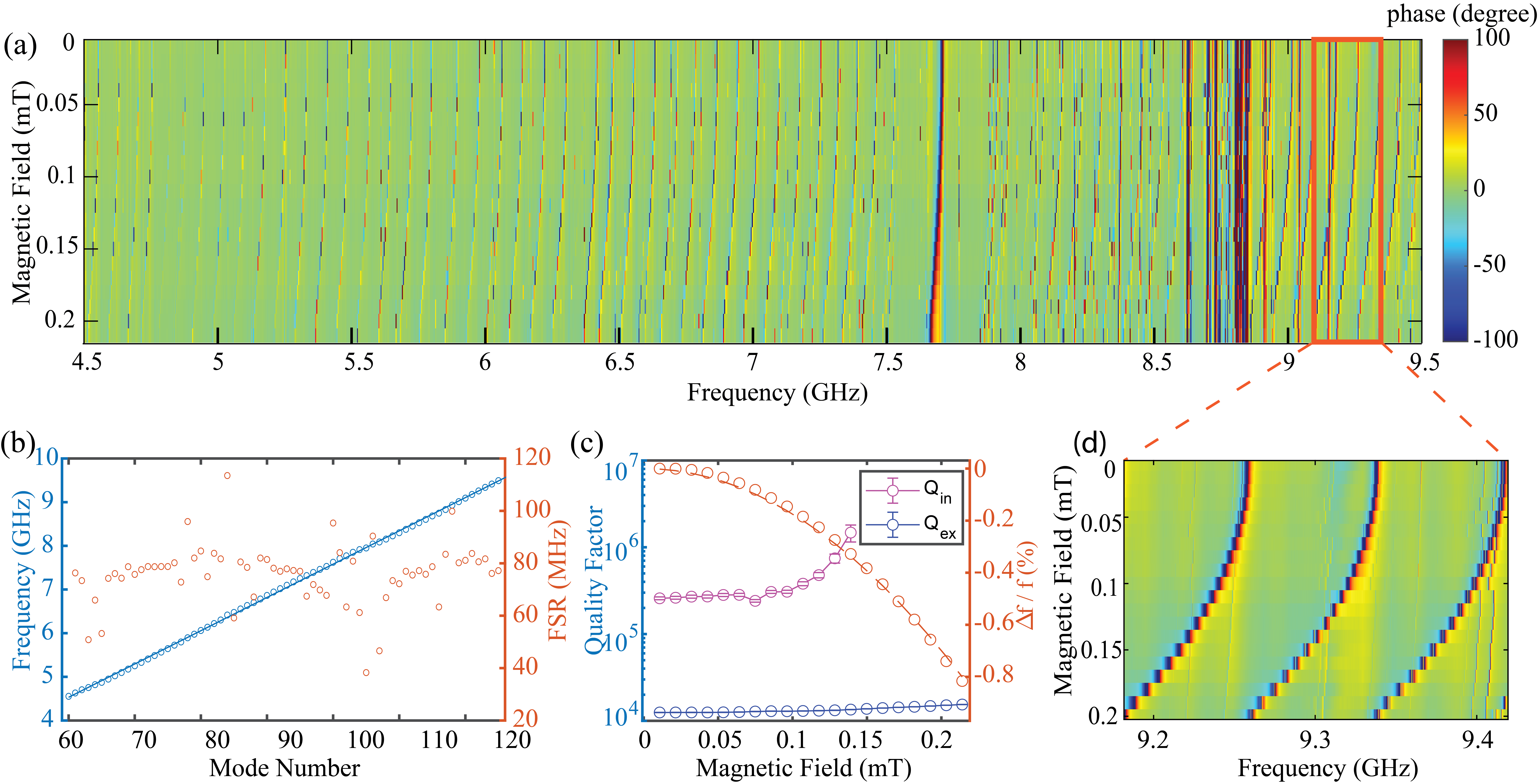}
    \caption{(a) Phase reflection spectrum of the resonator under tuned out-of-plane magnetic fields. The inset on the bottom right provides a zoomed-in view of the spectrum around 9.3\,GHz. (b) Left axis: blue dots indicate the mode frequency versus the mode number, with the line plot showing the linear fit of the frequency. The slope of the fit indicates an FSR of 76 MHz. Right axis: orange circles represent the free spectral range versus the mode number. (c) Intrinsic quality factor ($Q_{\mathrm{in}}$, magenta circle), external quality factor ($Q_{\mathrm{ex}}$, blue circle) and fractional modal frequency shift (orange circle) as a function of applied out-of-plane magnetic field shift, plotted with the fitted lines.
    }
    \label{fig:2-frequency_tuning}
\end{figure*}

Figure \ref{fig:2-frequency_tuning} (a) shows the reflection spectrum of the resonator under different out-of-plane magnetic fields. Figure \ref{fig:2-frequency_tuning} (b) displays the free spectral range (FSR) and mode frequency versus mode number. By performing a linear fit of the mode frequency, we determine the average FSR over the measured range to be approximately 76 MHz. Notably, the FSR in the upper panel exhibits a larger dispersion than expected from the design alone (see Appendix \ref{sec:dispersion} for a detailed analysis), which we attribute primarily to coupling with package modes.

\begin{equation}
    \frac{\Delta f}{f} = -\frac{\gamma}{2} \frac{d^2}{L_{\mathrm{dc}}^2 I_{2}^{*2}} B_\mathrm{ext}^2. 
    \label{equ:freq_shift}
\end{equation}
Figure~\ref{fig:2-frequency_tuning}(c) shows the frequency shift and quality factor change of a resonance mode at around 4.85 GHz under an applied magnetic field. The frequency shift is fitted using Eq. \ref{equ:freq_shift}, with a maximum shift of 0.83\%. This corresponds to a frequency tuning of 49\% of the FSR near 4.85 GHz and full FSR tuning near 9.40 GHz, as shown in Figure \ref{fig:2-frequency_tuning}(d). It is important to note that the limited frequency tuning range at lower frequencies can be overcome with the following methods. First, by making the two nanowires less asymmetric (i.e., $\gamma \rightarrow 1$), the frequency tuning ratio can be increased proportionally as indicated in Eq. (\ref{equ:freq_shift}). Second, by creating a longer ring to reduce the FSR, the same frequency tuning range can cover a greater proportion of the FSR. With these improvements, it is possible to achieve 100\% frequency tuning ratio between 4-8 GHz.

It is important to note that with the current tuning mechanism, the signal and idler frequencies cannot be tuned independently, i.e., when a current is applied, all modes shift proportionally to their zero-bias frequencies. Nevertheless, this device can effectively bridge the frequency gap (several Gigahertz) between two quantum information processing modules (denoting Module A and Module B), provided one of the following conditions is met: (1) Module A (matching the signal mode) or Module B (matching the idler mode) is frequency-tunable, or (2) Module A or B has a bandwidth larger than the converter’s FSR. Frequency tunability is a common feature in superconducting quantum interference devices (SQUIDs) based on Josephson junctions, with tuning ranges reaching up to hundreds of MHz \cite{houck2007generating}. In addition, large-bandwidth quantum information processing modules, such as microwave-to-optical converters with a conversion bandwidth of 20 MHz, have been demonstrated \cite{mckenna2020cryogenic}. To match such performance, the FSR of our converter would need to be reduced to a similar level, which can be achieved by increasing the total length of the structure or enhancing its kinetic inductance.

The quality factors, as shown in Fig.~\ref{fig:2-frequency_tuning}(c), remain approximately constant for magnetic fields below 0.1\,mT. However, at higher magnetic fields, the internal quality factor ($Q_{\mathrm{in}}$) exhibits an increasing trend. This is likely due to reduced coupling to the strongly-coupled two-level systems (TLSs) (see Appendix~\ref{sec:app_qWithMagFieldChange} for further details).

\section{Coherent Frequency Conversion}
\begin{figure*}[t]
    \centering
    \includegraphics[width=\textwidth]{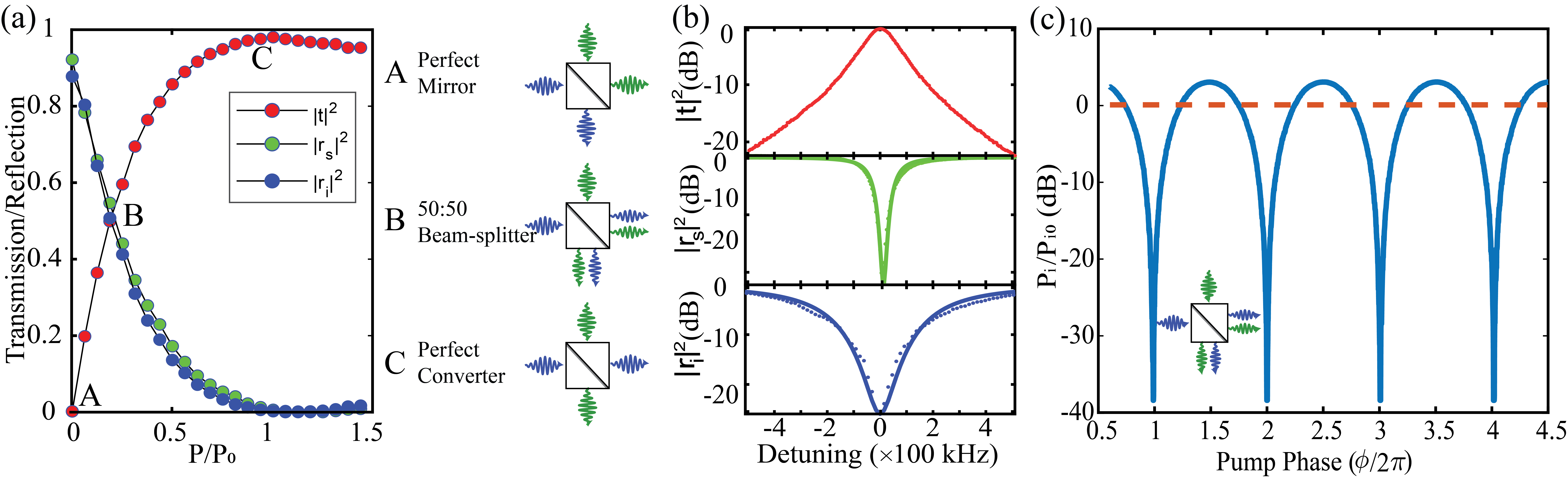}
    \caption{(a) The transmission and reflection spectra are shown as functions of the normalized pump power, where $P_0$ is the pump power that reaches the highest conversion efficiency. The signal transmission is marked by red dots, while signal and idler reflections are indicated by green and blue dots, respectively. Labels A, B, and C denote three distinct pump conditions, with an accompanying beam-splitter function illustrated on the right. (b) The transmission and reflection spectra are depicted with dotted lines, accompanied by a Lorentzian fit shown as solid lines. (c) The coherent interference fringes of the reflected idler power are displayed as a function of the pump phase. The inset illustrates the beam-splitter process, where both signal and idler probes, each with the same photon number, are directed into the device.}
    \label{fig:3_transduction}
\end{figure*}

In principle, one can pick two arbitrary modes for frequency conversion. We refer to the lower frequency mode as the signal and the higher frequency mode as the idler. When an out-of-plane magnetic field applied, the resulting dc current introduces a $\chi^{(2)}$ nonlinearity into a $\chi^{(3)}$ medium. This allows for frequency conversion with a single pump tone, which drives at the difference frequency between the signal and idler mode. In a dispersive multimode resonator, where the spectral spacing between adjacent modes varies, this can be advantageous for mode selection. The non-uniform spacing allows for pumping at the frequency difference between two specific modes without introducing coupling to any other modes. In this scenario, the scattering parameters can be related to the cooperativity ($C$) of the two modes as 
\begin{equation}
\begin{split}
    |t|^2 & = \eta_\mathrm{s} \eta_\mathrm{i} \frac{4C}{(1+C)^2}, \\
    |r|^2 & = 1 - \eta_\mathrm{s} \eta_\mathrm{i} \frac{4C}{(1+C)^2}, \\
    \label{equ:transmission_reflection}
\end{split}
\end{equation}
where $t$ and $r$ are the on-chip transmission and reflection coefficients, and $|t|^2$ is also referred to as the conversion efficiency in the context of parametric frequency transduction. The coupling efficiency is defined as the ratio between the external coupling rate and total coupling rate, $\eta_{\mathrm{s(i)}} = {\kappa_{\mathrm{s(i), ex}}}/{\kappa_{\mathrm{s(i)}}}$. The cooperativity can be expressed as 
\begin{equation}
    C = \frac{4g_0^2 n_\mathrm{eff}}{\kappa_\mathrm{s} \kappa_\mathrm{i}},
\end{equation}
where $g_0$ is the single-photon nonlinear coupling rate between the two modes, and $n_\mathrm{eff}$ is the effective pump photon number. Here the cooperativity equals to the normalized pump power $P/P_0$, where $P_0$ is the pump power that achieves highest conversion efficiency.  From Eq. (\ref{equ:transmission_reflection}), it can be seen that the conversion efficiency is upper bounded by $\kappa_{\mathrm{s, ex}} \kappa_{\mathrm{i, ex}}/{\kappa_{\mathrm{s}} \kappa_{\mathrm{i}}}$. This implies that to achieve unity transmission, both the signal and idler modes must be strongly overcoupled. For our device, the average internal quality factor is $3.93\times 10^5$ (at the probe photon number of $10^5$), benefiting from the large mode volume. The average coupling efficiency is 94\%, indicating that it is an ideal platform for frequency conversion.

To demonstrate noiseless frequency conversion, we apply an out-of-plane magnetic field of 0.14~mT and select two resonant modes with frequencies \(f_s = 4.85~\mathrm{GHz}\) and \(f_i = 6.61~\mathrm{GHz}\). A pump tone, set to the frequency difference between the two modes (\(f_p = f_i - f_s\)), is introduced through the input port alongside a weak signal probe tone. To mitigate the heating introduced by the pump tone, we send a pulsed pump with a duty cycle of 1/100, and pulse duration of 10\,$\mu$s. An idler tone is generated through the frequency-conversion process. The signal, idler, and pump tones are then reflected from the input port. A circulator redirects these reflected waves to the output line, where the pump tone is filtered out using a high-pass filter. The output microwaves are subsequently amplified by a high-electron-mobility transistor (HEMT) amplifier located on the 4-K plate of the dilution refrigerator. To characterize the conversion performance, we adopt the bidirectional conversion measurement approach outlined in Ref. \cite{xu2021bidirectional}. The on-chip conversion efficiency is determined as follows
\begin{equation}
    |t|^2 = \frac{S_{\mathrm{is, p}} S_{\mathrm{si, p}}}{S_{\mathrm{ss, bg}} S_{\mathrm{ii, bg}}},
\end{equation}
where $S_{\mathrm{is, p}}$ and $S_{\mathrm{si, p}}$ represent the peak magnitudes in the transmission spectrum, and $S_{\mathrm{ss, bg}}$ and $S_{\mathrm{ii, bg}}$ represent the background magnitudes in the reflection spectrum. This method allows us to compensate for off-chip gain and loss factors affecting each mode. 

\begin{figure*}[t]
    \centering
    \includegraphics[width=\textwidth]{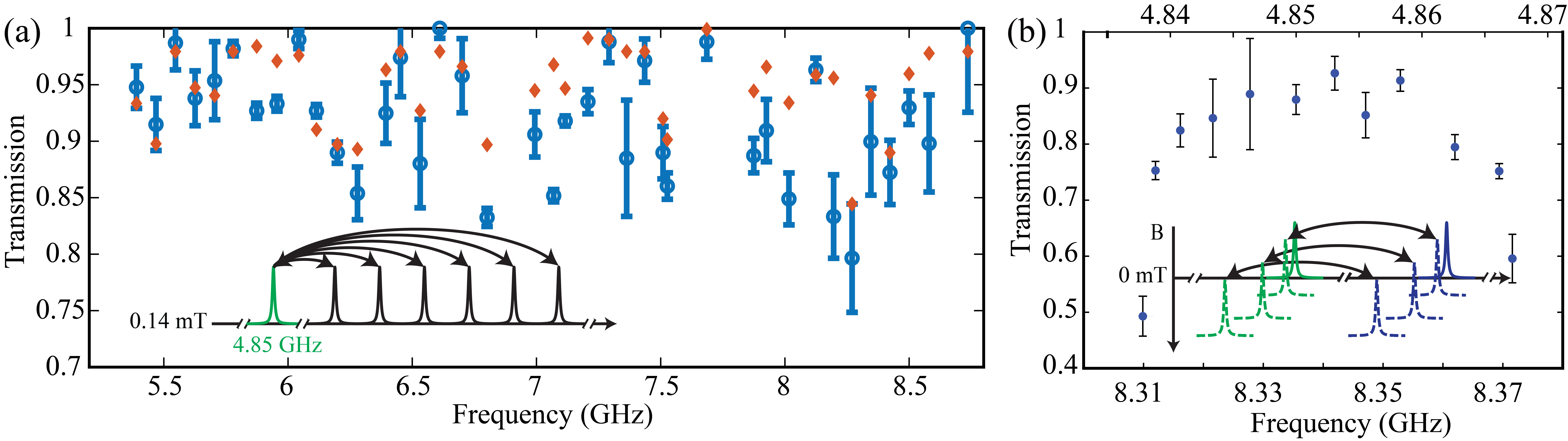}
    \caption{(a) Conversion efficiency, indicated by blue circles with error bars, measures with a fixed signal mode while varying the idler modes. The inset illustrates the process, highlighting the signal mode in green. (b) Conversion efficiency, depicted as blue circles with error bars, measured with a fixed pair of signal and idler modes while the frequency is tuned by an out-of-plane magnetic field. The top x-axis displays the signal frequency, and the bottom x-axis shows the idler frequency. The inset depicts the frequency shift of a signal mode (green) and an idler mode (blue) under an increasing magnetic field.}
    \label{fig:4-broadbandConversion}
\end{figure*}

In Fig.~\ref{fig:3_transduction}(a), we illustrate measurements of the reflection parameters at the signal mode (green dots) and the idler mode (blue dots), as well as the conversion parameters, plotted as functions of the cooperativity. The conversion parameter increases with cooperativity, reaching its maximum when cooperativity is at unity. This conversion process mimics the beam-splitter interaction, where the conversion between the signal and idler is analogous to the transmission through a beam-splitter. The process unfolds in three distinct stages, labeled A, B, and C, shown in the inset. Stage A, with zero cooperativity, reflects the signal completely, functioning as a perfect mirror. Stage B, where cooperativity equals to $3-2\sqrt{2}$, exhibits an equal balance of transmission and reflection. Finally, stage C is characterized by full transmission when cooperativity reaches one. Here the maximum transmission coefficient is $98.5 \% \pm 0.8 \%$, which is limited by the coupling efficiency of the signal and the idler modes. The signal and idler coupling efficiency are measured $\eta_s = 0.99 \pm 0.01$, $\eta_i = 0.98 \pm 0.01$. The internal conversion efficiency is inferred to be almost ideal. The transmission and reflection spectrum at unity cooperativity is shown in Fig.~\ref{fig:3_transduction}~(b) as a function of the probe detuning. The magnitude of the reflection coefficient is below -20 dB, indicating a well phase-matched condition. The conversion bandwidth, which is limited by both the signal and idler mode bandwidth, is around $\Gamma/2\pi = 130$~kHz. 

An ideal frequency converter should also be noiseless. To characterize the added noise of the conversion process, we use a variable temperature stage (VTS) to introduce known thermal noise to the device's input port \cite{xu2023magnetic, xu2024radiatively}. The thermal occupancy of the signal and idler ports is measured as 0.59 $\pm$ 0.04 quanta and 0.79 $\pm$ 0.08 quanta, respectively (see Appendix \ref{sec:app_noiseCalibration} for details on noise calibration and analysis). The elevated thermal occupancy is due to the strong pump power required for the weak nonlinearity $K=2\pi \times0.1$ Hz, estimated from the bifurcation power \cite{eichler2014controlling}. One possible solution to mitigate this issue is to fabricate thinner and narrower nanowires, which would increase nonlinearity and consequently reduce the pump power requirement. 

To demonstrate coherent frequency conversion, in which the up-converted idler tone acquires a phase shift that depends on the pump phase, we operate the device as a 50:50 beam-splitter. In addition to the pump tone, we introduce both the signal and idler tones with equal photon numbers and fixed phase into the input port. By sweeping the phase of the pump tone, the transmitted idler tone from the signal obtains a corresponding phase shift, resulting in coherent interference with the reflected idler tone, as shown in Fig.~\ref{fig:3_transduction}~(c). The solid blue line represents the normalized idler power $P_i/P_{i0}$ as a function of the pump phase, where $P_{i0}$ is the input idler power. The orange dashed line represents the reflected idler power without the pump. In the case of constructive interference, the output idler power is 3 dB above the reference value. In the case of destructive interference, we observe transmission to be as low as -40 dB. This is equivalent to an interference visibility of 99.99\% of the input signal and idler beams. 

Thus far, we have demonstrated low-noise and near-unity conversion efficiency between two specific resonant modes. Next, we show that it is possible to achieve similar results between any two resonant modes of the device. To simplify the measurement, we fix the signal mode and perform frequency conversion with different idler modes, as shown in the inset of Fig. \ref{fig:4-broadbandConversion}(a). The probe power is fixed at -100 dBm. The blue circles with error bars represent the measured conversion efficiency, while the orange diamonds indicate the total coupling efficiency, $\eta_s \eta_i$, which is the upper bound for the conversion efficiency. The average conversion efficiency is 91\% over the entire measured frequency span. All measured conversion efficiencies across the span exceed 80\%, highlighting this as a promising multimode platform for efficiently coupling any two modes. The high efficiency benefits from the low internal loss of the device. To further enhance the overall conversion efficiency, the resonator can be designed with an enhanced external coupling rate. With a tenfold increase in the external coupling rate, we expect the average conversion efficiency to reach 99\%.

To establish the broad spectrum coverage, it is also important to demonstrate the frequency conversion with mode frequency tuned by the magnetic field, as illustrated in the inset of Fig. \ref{fig:4-broadbandConversion}(b). Here, we choose the idler modes at around 8.3\,GHz, while keeping the signal mode the same. We measure the conversion efficiency while tuning the out-of-plane magnetic field, as shown in Fig. \ref{fig:4-broadbandConversion}(b). With the increase of the magnetic field, the frequency starts to shift to the lower frequency, while the conversion efficiency increases. This aligns with the increase of the internal quality factor observed in Fig. \ref{fig:2-frequency_tuning} (c). The conversion efficiency peaks at around 4.853~Ghz with around 92\%. Upon further increasing the magnetic field strength, the efficiency drops down to 49\%. This is likely due to the presence of higher-order nonlinearity when the dc current approaches the characterization current. 

The multimode resonator also benefits the pump delivery. Because it supports multiple modes, the pump frequency—defined as the difference between the signal and idler frequencies—coincides (with slight detuning due to dispersion) with another resonant mode. This additional mode effectively enhances the pump tone, a feature not available in devices like the Josephson Parametric Converter (JPC), which hosts only two resonant modes.


\begin{figure*}[t]
    \centering
    \includegraphics[width=0.7\linewidth]{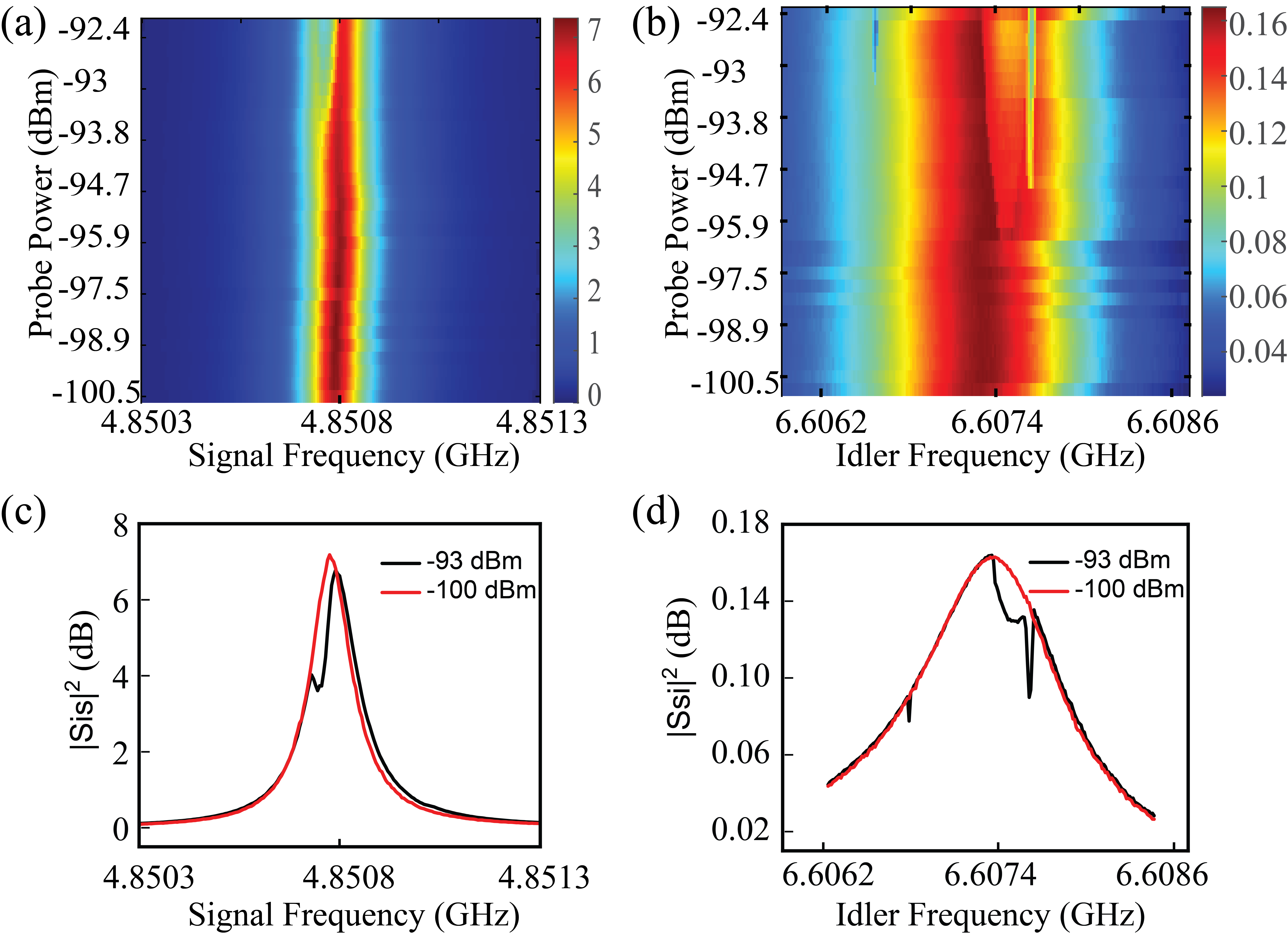}
    \caption{(a-b) 2D heat plot of the transmission spectrum of the signal (a) and the idler (b) as a function of the probe power. The asymmetric spectrum with dips at a higher power indicates the conversion has been saturated. (c-d) display the cross-section of the transmission spectrum at probe powers of -93 dBm and -100 dBm, respectively, providing a clear illustration of the spectrum when the probe power exceeds the bifurcation power.}
    \label{fig:5-saturation_power}
\end{figure*}

The maximum input power that the device can handle is limited by the bifurcation power of the resonator. Figure \ref{fig:5-saturation_power} (a-b) shows a 2D heat plot of the transmission spectrum as a function of probe power and signal (idler) frequency. The transmission spectrum maintains a Lorentzian shape when the probe power is below the bifurcation power ($\approx$ -95\,dBm). At a higher probe power, the transmission spectrum begins to deviate from an ideal Lorentzian shape. Figures \ref{fig:5-saturation_power} (c-d) illustrate this deviation by comparing the transmission spectrum of the signal and idler frequencies at -100 dBm and -93 dBm, respectively. The saturation power is similar to that of a Josephson-junction-based device \cite{abdo2013full}. In addition to the Kerr nonlinearity-induced saturation, we also observed shape dips in the conversion spectrum, as shown in Fig. \ref{fig:5-saturation_power}(d). We attribute these dips to localized breaks in superconductivity caused by device defects \cite{romanenko2014dependence}.

The device is also capable of converting signals at the single-photon level. One common challenge at lower signal power is the degraded internal quality factor due to coupling with two-level systems (TLSs) \cite{pappas2011two}. In our case, the average internal quality factor at the single-photon level decreases to the order of $10^4$, comparable to the coupling quality factor. This would upper bound the conversion efficiency to 25\%. However, this issue can be mitigated by applying a detuned microwave drive near both the signal and idler modes to saturate the two-level system \cite{andersson2021acoustic} without affecting the conversion process. By adopting this approach, we improve the conversion efficiency from 26\% to 93\% at a single-photon level (see Appendix \ref{sec:app_SinglePhotonConversion} for detailed results). These advances highlight the device's potential for high-efficiency quantum information processing at the single-photon level. 

\section{Discussion}
In conclusion, we have presented a multimode kinetic inductance metastructure wherein each resonant mode can be tuned by an external magnetic field. We demonstrated that it is possible to achieve efficient microwave-to-microwave (M2M) conversion between any two selected resonances near their ground state. This capability enables the device to function as a quantum transducer, connecting any two microwave modes between 4.85 GHz to 8.5 GHz. It is important to note that the frequency conversion range demonstrated here is mostly limited by our measurement bandwidth of the signal readout chain configured in the dilution refrigerator. In principle, this device can operate at a higher frequency up to the millimeter-wave range \cite{anferov2020millimeter}, making it more attractive than the Josephson junction counterparts.

The device can also function as a parametric amplifier \cite{parker2022degenerate, xu2023magnetic, mohamed2024selective, khalifa2024kinetic} with a wide frequency tuning range (see Appendix \ref{sec:app_amplification} for demonstration of parametric amplification using this device). The three-wave mixing Hamiltonian induced by an applied magnetic field, enables degenerate parametric amplification with a single-tone pump that can be effectively filtered. Additionally, it is possible to generate a two-mode squeezed state over any pair of resonances based on the parametric down-conversion process \cite{wu2024junction}. Benefiting from its relatively high critical temperature, such a device can also operate at elevated temperatures \cite{xu2024radiatively, frasca2024three}.

Such a device can be useful in many applications. Firstly, it can be combined with an M-O conversion device to bridge any microwave mode to telecom light. Additionally, it can extend the microwave frequency coverage of a quantum information processor, such as microwave single-photon sources \cite{houck2007generating} and microwave single-photon detectors \cite{wang2023single}. This device can also mediate coupling between two spatially and temporally separated modes. For instance, by flip-chip bonding this device with a bulk-acoustic resonator, it is possible to couple any two spectrally separated acoustic modes \cite{han2016multimode, han2022superconducting} and magnonic modes \cite{zhang2016optomagnonic, zhang2016cavity}. A similar application can be found in routing interactions between two quantum information processors. Ultimately, this versatile device paves the way for more efficient and integrated quantum networks and information processing systems in the microwave regime.

\begin{acknowledgments}
This project is supported by the US Department of Energy Co-design Center for Quantum Advantage (C2QA) under Contract No. DE-SC0012704 and in part by the Defense Advanced Research Projects Agency under cooperative agreement HR0011-24-2-0346. YFW and HXT acknowledge the support from the Office of Naval Research under contract number N00014-23-1-2121. The authors would like to thank Professor Liang Jiang and Ming Yuan, for useful discussions. We thank Dr. Yong Sun, Dr. Lauren McCabe, Mr. Kelly Woods, and Dr. Michael Rooks for assistance in device fabrication. \end{acknowledgments}

\appendix

\section{Resonant Mode Analysis}
\label{sec:app_modeAnalysis}
\begin{figure}[t]
    \centering
    \includegraphics[width=\linewidth]{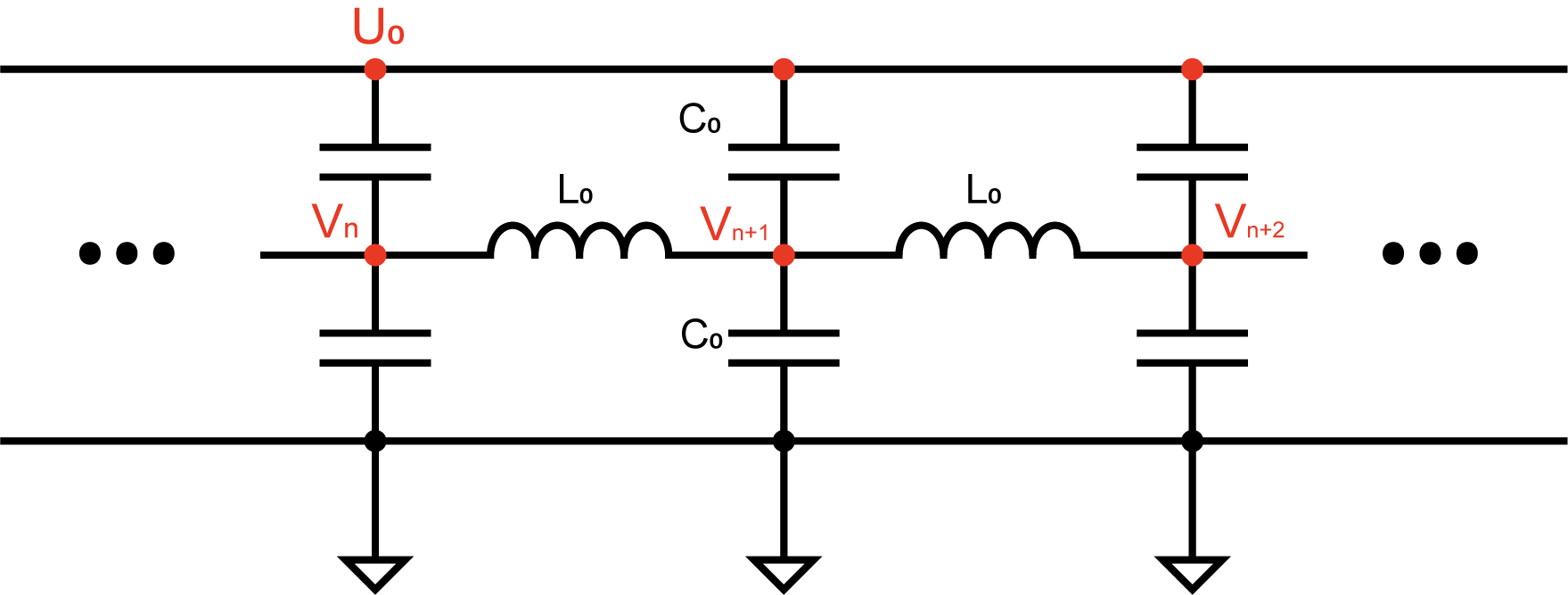}
    \caption{Equivalent circuit diagram for modal analysis. The diagram shows a segment of the main device structure. }
    \label{fig:eqv_circuit}
\end{figure}

\begin{figure*}
    \centering
    \includegraphics[width=0.9\linewidth]{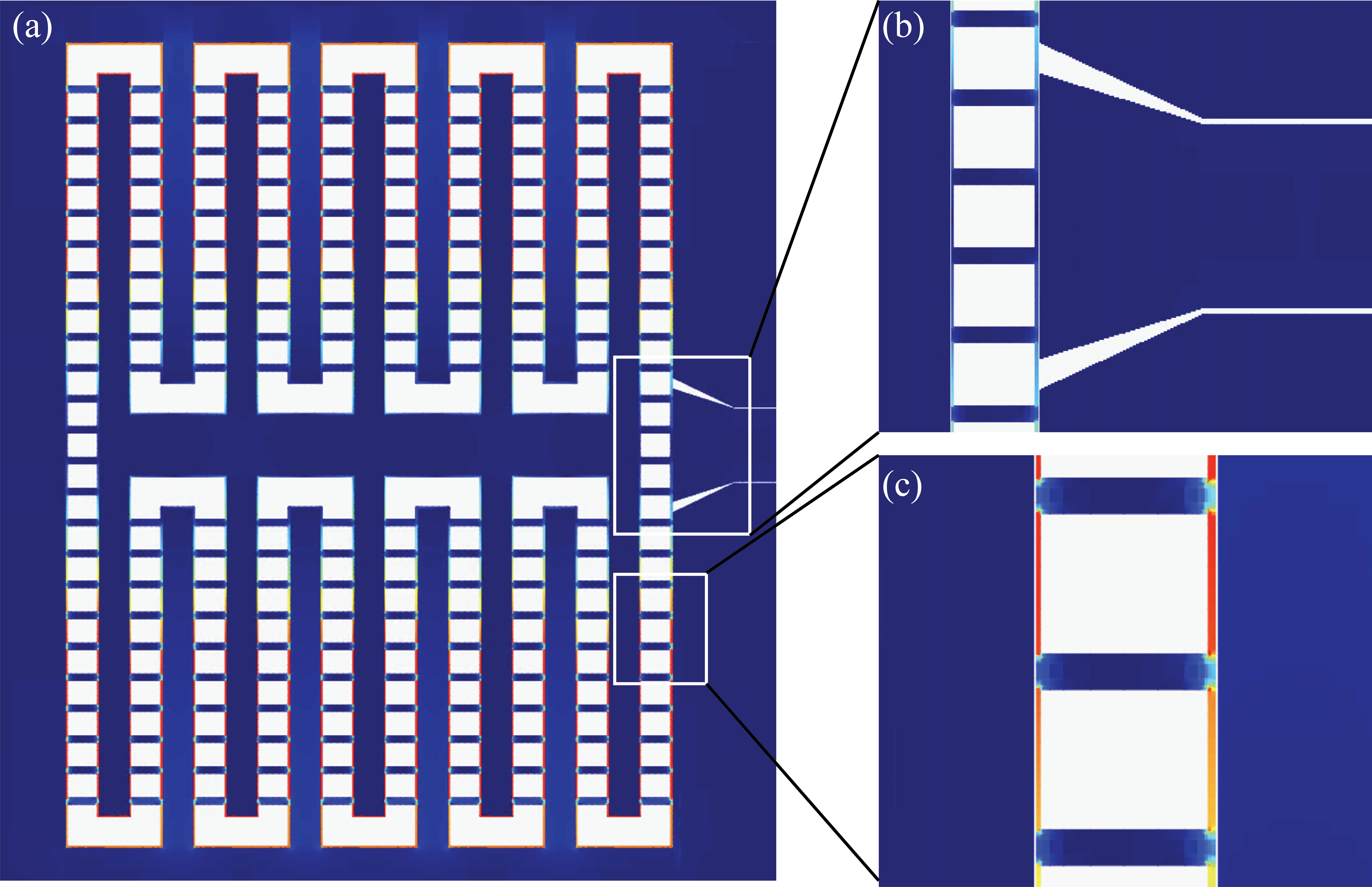}
    \caption{(a) Current density distribution of the fifth resonant mode in an effective pattern with reduced length. The color scale represents the magnitude of the current density, with red indicating high intensity and blue indicating low intensity. (b) Enlarged view of the current distribution near the coupling port. (c) Enlarged view of the current distribution inside a single microloop.}
    \label{fig:simulation}
\end{figure*}

In this section, we analyze the resonant frequencies of our circuit under zero bias current. The analysis is based on the equivalent circuit depicted in Fig.~\ref{fig:eqv_circuit} to determine the resonant modes. In this model, each microloop, as shown in Fig.~\ref{fig:1-device}(b), is represented as a lumped-element LC circuit. The effective inductance and capacitance are given by \( L_0 = (L_k + L_m)l_0 \) and \( C_0 = \frac{1}{2}Cl_0 \) respectively, where \( L_k \) and \( L_m \) denote the kinetic and geometric inductances per unit length, \( C \) is the capacitance per unit length, and \( l_0 \) is the length of a single loop. This modeling approach is valid for modes whose indices are significantly smaller than the total number of microloops, \( N = 3200 \), ensuring that the phase shift in each loop is substantially less than \( 2\pi \).

The above treatment allows us to treat the circuit in a similar fashion as a uniform transmission line with a notable distinction: one side of the line connects to an isolated island with potential $U_0$, as depicted in Fig. \ref{fig:eqv_circuit}, instead of to a ground As we will demonstrate, this configuration induces a static shift in the voltage amplitude but does not influence the resonant frequencies of the circuit.

In light of the previous discussion, we establish the constraints for the system in a time-harmonic steady state. Applying Kirchhoff's laws to the circuit elements yields
\begin{align}
    \frac{V_{n+1}-V_n}{i\omega L_0}+\frac{V_{n+1}-V_{n+2}}{i\omega L_0}+i\omega C_0(2V_{n+1}-U_0)=0,
\end{align}
where $V_n$ is the voltage at the n-th microloop. Rewriting the above equation as 
\begin{align}
    V_n+V_{n+2}-2V_{n+1}+\frac{\omega^2}{\omega_0^2}(2V_{n+1}-U_0)=0,
    \label{equ:VEquation}
\end{align}
where $\omega_0=1/\sqrt{L_0C_0}$ is the natural angular frequency of the LC circuit. To solve Eq. \ref{equ:VEquation}, we introduce the travelling-wave ansatz 
\begin{align}
    V_n=e^{ikl_{0}n}A+B,
    \label{eq:ansatz}
\end{align}
where $A$ and $B$ are constants, and $k$ is the wave number. Substituting the ansatz Eq. \ref{eq:ansatz} into Eq. \ref{equ:VEquation} and simplifying, we obtain 
\begin{align}
    [\frac{\omega^2}{\omega_0^2}-(1-\cos{kl_0})]e^{ikl_0(n+1)}A+\frac{\omega^2}{\omega_0^2}(B-2U_0)=0.
\end{align}
For this equation to apply for arbitrary $A$ and $n$, we require the coefficient of $e^{ikl_0(n+1)}$ to be zero
\begin{align}
    \frac{\omega^2}{\omega_0^2}-(1-\cos{kl_0})=0.
    \label{eq:omega1}
\end{align}
The periodicity of the structure requires 
\begin{align}
    V_{N+1}=V_1\Rightarrow kl_0N=2\pi\times m,
\end{align}
Thus, $k$ in Eq. \ref{eq:ansatz} can only take a series of discrete values 
\begin{align}
    k_m=\frac{2\pi}{l_0N}m,
\end{align}
where $m\in\{0,..,N-1\}$. This, combined with Eq. \ref{eq:omega1}, determines the system's resonant frequencies, which can use $m$ as the mode index 
\begin{align}
    \omega_m=\omega_0\sqrt{1-\cos{k_ml_0}}\approx\frac{\sqrt{2}\pi m}{N}\omega_0.
\end{align}
Here we used the fact that this approximated treatment only applied when the phase accumulated across a single loop is negligible compared to $2\pi$, i.e. $k\ll 2\pi$.

Furthermore, using the fact that 
\begin{align}
    \frac{\omega_0}{N}=\sqrt{\frac{2}{(L_k+L_m)C}}\frac{1}{l},
\end{align}
where $l=Nl_0$ is the total length of the device, we obtain 
\begin{align}
    f_m=\sqrt{\frac{1}{(L_k+L_m)C}}\frac{m}{l}.
\end{align}
Thus, the free spectrum range of the device is 
\begin{align}
    \text{FSR}=\sqrt{\frac{1}{(L_k+L_m)C}}\frac{1}{l}.
\end{align}


To derive the theoretically predicted FSR, we conducted a finite element analysis using a reduced total length effective pattern in Sonnet $em$ microwave simulation. Shown in Fig. \ref{fig:simulation} (a) is a mode simulation of a simplified structure with a total length of 5.7 mm. The plotted mesh is the current distribution of the fifth resonant mode at 5.635~GHz, with the node and the antinode shown as high (blue) and low (red) current density. Fig. \ref{fig:simulation} (b) and (c) show the enlarged current distribution near the coupling port and in a microloop. 

The kinetic inductance of the thin NbN film is characterized as $L_{k}^{\square} = 105\, \mathrm{pH/\square}$, corresponded to $L_k = 57\, \mu\mathrm{H/m}$ for the meta-ring. Other parameters of the meta-ring is obtained by Sonnet $em$ microwave simulation: $L_m = 0.25 \, \mu\mathrm{H/m}$, the characteristic impedance is $Z_c = \sqrt{{(L_k+L_m)}/{C}}=362 \, \mathrm{\Omega}$, and the phase velocity is $v_{\mathrm{ph}} = 1/\sqrt{{(L_k+L_m)}{C}}=6.3\times 10^{6}\,\mathrm{m/s}$. These values predict a FSR of approximately 79 MHz for a $80~\mathrm{mm}$ long meta-ring, which closely matches the experimentally measured FSR of $76~\mathrm{MHz}$.


\section{Dispersion Analysis}
\label{sec:dispersion}

\begin{figure*}
    \centering
    \includegraphics[width=\linewidth]{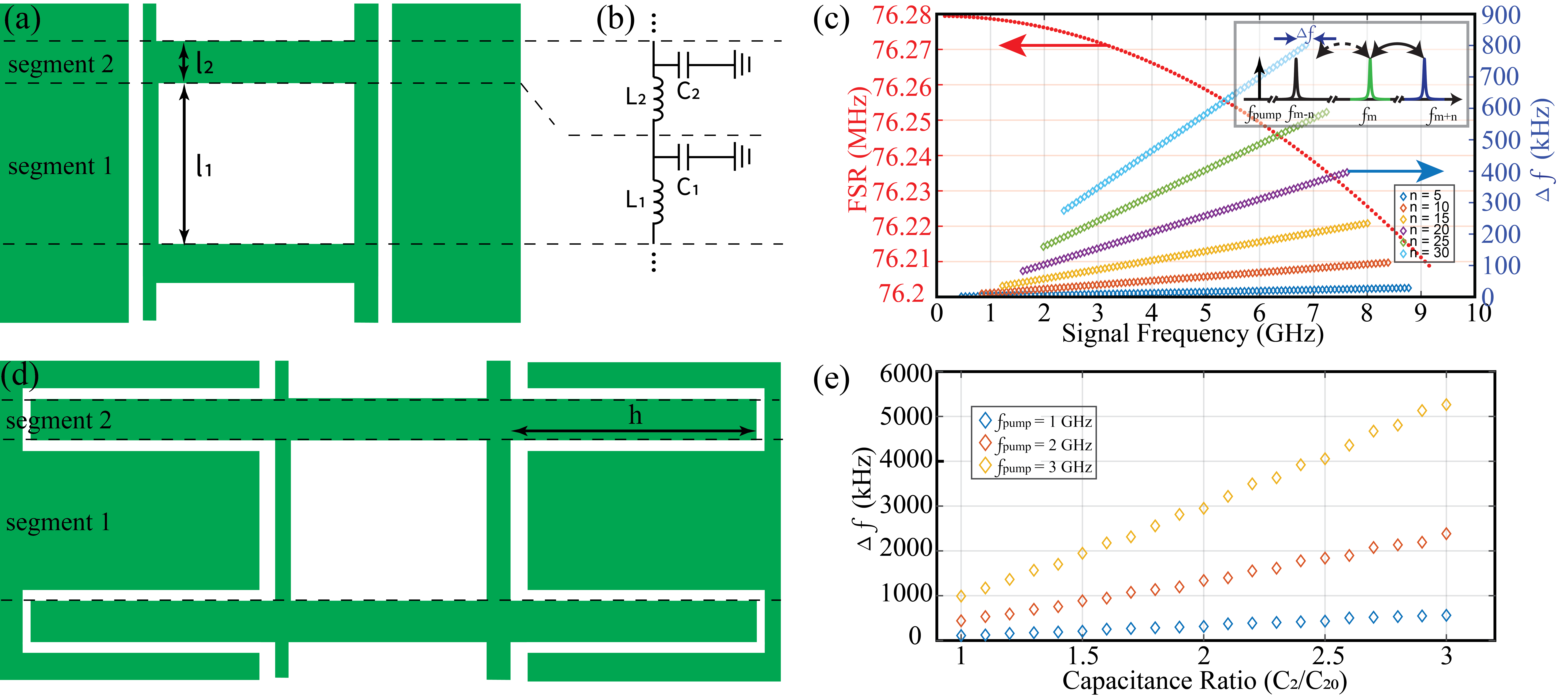}
    \caption{(a) Schematic of a microloop, divided into two segments as indicated by the dashed lines.
(b) Equivalent circuit representation of the microloop structure.
(c) Left axis: Free spectral range (FSR) of the current device design as a function of the signal frequency.
Right axis: Frequency mismatch between $f_m$ and $f_{m-n}$ when driving the conversion between modes $m$ and $m+n$. The inset illustrates the conversion process.
(d) Proposed design featuring an interdigitated capacitor structure. The capacitance can be adjusted by varying the finger length $h$.
(e) Frequency mismatch during the conversion process between a signal mode at 5 GHz and idler modes that are 1, 2, and 3 GHz apart, plotted as a function of the capacitance enhancement ratio in segment 2.}
    \label{fig:app_dispersion}
\end{figure*}

In Appendix \ref{sec:app_modeAnalysis}, we analyzed the mode frequency under the non-dispersion assumption. In this section, we consider the dispersion induced by the varying inductance and capacitance from the microloop structure. 

We break the circuit diagram corresponding to a single unit cell, as shown in Fig.~\ref{sec:app_modeAnalysis}, into two segments: one corresponding to the parallel nanowires (segment 1) and the other to the bridge connecting the nanowires (segment 2), as illustrated in Fig.~\ref{fig:app_dispersion}(a). Each segment has a distinct per-unit-length inductance and capacitance, as indicated in Fig.~\ref{fig:app_dispersion}(b). The length of the unit cell is given by the sum of the lengths of the two segments$l_0 = l_1 + l_2$.

The dispersion relation can be derived by writing the ABCD matrix for the two segments \cite{malnou2021three}:
\begin{equation}
\begin{aligned}
    M_1(\omega) &=
    \begin{pmatrix}
        \cos[k_1(\omega) l_1] & jZ_1 \sin[k_1(\omega) l_1] \\
        j \tfrac{1}{Z_1} \sin[k_1(\omega) l_1] & \cos[k_1(\omega) l_1]
    \end{pmatrix},\\[8pt]
    M_2(\omega) &=
    \begin{pmatrix}
        \cos[k_2(\omega) l_2] & jZ_2 \sin[k_2(\omega) l_2] \\
        j \tfrac{1}{Z_2} \sin[k_2(\omega) l_2] & \cos[k_2(\omega) l_2]
    \end{pmatrix},
\end{aligned}
\end{equation}
where $Z_j = \sqrt{L_j / C_j}$ represents the characteristic impedance, and $k_j$ is the wave number for the $j-th$ segament. The unit cell matrix is obtained by multiplying the two matrices:
\begin{equation}
    M_{\text{cell}}(\omega) = M_2(\omega) M_1(\omega) =
    \begin{pmatrix}
        A_{\text{cell}}(\omega) & B_{\text{cell}}(\omega) \\
        C_{\text{cell}}(\omega) & D_{\text{cell}}(\omega)
    \end{pmatrix}.
\end{equation}
In an infinitely repeated structure with period $l_0$, the electromagnetic waves must satisfy the Bloch condition:
\begin{equation}
    \Psi(x + l_0) = e^{j k l_0} \Psi(x),
\end{equation}
where $k$ is the Bloch wave number. It follows from standard network theory (or chain-matrix arguments) that
\begin{equation}
    \mathrm{Tr}(M_{\text{cell}}(\omega)) = A_{\text{cell}}(\omega) + D_{\text{cell}}(\omega) = 2 \cos(k l_0),
\end{equation}
which can be rewritten as
\begin{equation}
\begin{aligned}
    \cos(k l_0) & = \cos(k_1 l_1) \cos(k_2 l_2) \\
                & - \frac{1}{2} \left( \frac{Z_1}{Z_2} + \frac{Z_2}{Z_1} \right) \sin(k_1 l_1) \sin(k_2 l_2).
\end{aligned}
\end{equation}
The periodic condition gives $\cos(k l_0) = \cos(2 \pi m / N)$, leading to the final dispersion relation:
\begin{equation}
\begin{aligned}
    \cos\left(\frac{2 \pi m}{N} \right) & = \cos(k_1 l_1) \cos(k_2 l_2) \\
                & - \frac{1}{2} \left( \frac{Z_1}{Z_2} + \frac{Z_2}{Z_1} \right) \sin(k_1 l_1) \sin(k_2 l_2).
\end{aligned}
\end{equation}
In our case, we use $l1 = 25 \mu\mathrm{m}$, $l2 = 5 \mu\mathrm{m}$, $L_1 = 57 \mu \mathrm{H/m}$, $L_2 = 3 \mu \mathrm{H/m}$, $C_1 = 289 \mathrm{pF/m}$, $C_2 = 880 \mathrm{pF/m}$. This gives the characterization impedance $Z_1 = 443~\Omega$ and $Z_2 = 71~\Omega$. In Fig. \ref{fig:app_dispersion} (c) (left axis), the FSR is calculated as a function of mode frequency. The FSR shows a frequency dependent decrease, indicate a dispersion feature as a result of the two-segment structure. This designed dispersion, however, is much smaller compared to what we actually measured as shown in Fig. \ref{fig:2-frequency_tuning} (c). One possible explanation for this discrepancy is the coupling between the resonator modes and parasitic package modes \cite{huang2021microwave}. Such coupling can lead to mode hybridization, causing irregular variations in the FSR rather than the expected monotonic decrease. 

We analyze how the two-segment induced dispersion enhances the conversion process by considering the transition of a signal mode (mode number \( m \)) to an idler mode (mode number \( m+n \)). A pump tone is applied at a frequency of \( f_{\mathrm{pump}} = f_{m+n} - f_{m} \). To maintain high conversion efficiency, the frequency difference \( f_{m} - f_{m-n} \) should deviate from the pump frequency by several resonance linewidths. The right axis of Fig.~\ref{fig:app_dispersion}(c) illustrates this frequency difference, defined as 
$\Delta f = 2 f_{m} - (f_{m+n} + f_{m-n})$, as a function of the signal frequency. The mode number difference \( n \) is varied from 5 to 30, corresponding to a signal-idler frequency separation ranging from 400 MHz to 2.4 GHz. At 5 GHz, \( \Delta f \) is 16 kHz for \( n=5 \) and increases to 586 kHz for \( n=30 \). Given an average resonance linewidth of 130 kHz, ensuring \( \Delta f \) exceeds three times the linewidth requires a signal-idler frequency difference of at least 2 GHz in the current dispersion design.

To enhance dispersion, one effective strategy is to implement an interdigitated capacitor (IDC) design \cite{malnou2021three, parker2022degenerate}, as illustrated in Fig.~\ref{fig:app_dispersion}(d). The length of the capacitor fingers, denoted as \( h \), can be adjusted to increase the capacitance in segment 2. In the following analysis, we assume the inductance remains unchanged in the new design. However, in practice, the inductance will decrease due to a reduced cross-sectional area, which would further enhance dispersion. Here, we focus solely on the effect of increased capacitance on dispersion, while the precise relationship between \( h \) and capacitance can be determined through electromagnetic simulations. Fig.~\ref{fig:app_dispersion}(e) examines the conversion of a signal mode at approximately 5 GHz to an idler mode located 1, 2, and 3 GHz away. The frequency mismatch, \( \Delta f \), is plotted as a function of the capacitance enhancement ratio \( C'_2/C_2 \), where \( C'_2 \) represents the enhanced capacitance. The results indicate that \( \Delta f \) increases nearly linearly with the capacitance enhancement ratio. For a capacitance enhancement factor of 3, \( \Delta f \) reaches 562~\text{kHz}, 2.383~\text{MHz}, and 5.263~\text{MHz} for idler frequencies 1~\text{GHz}, 2~\text{GHz}, and 3~\text{GHz} apart from the signal frequency, respectively. 

\section{Frequency Tuning and Hamiltonian Analysis}
\label{sec:app_freq_tuning}
\begin{figure}
    \centering
    \includegraphics[width=0.8\linewidth]{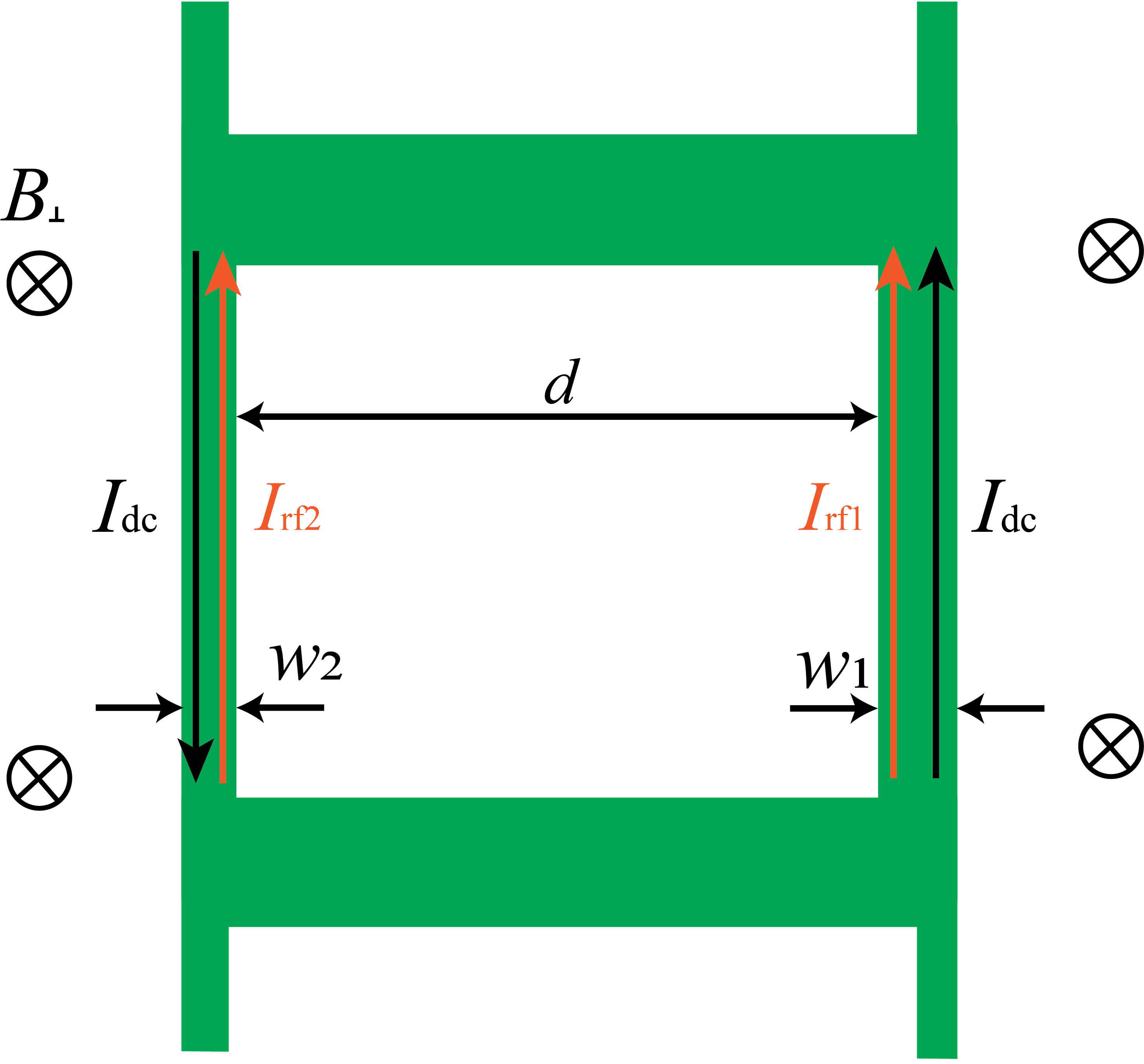}
    \caption{Illustration of the current distribution in a single microloop with asymmetric width. The inward magnetic field induces a DC supercurrent, depicted by the black arrowed lines, which circulates around the microloop in a counterclockwise direction. The RF currents in the two nanowires, shown with orange arrowed lines, flow in the same direction. }
    \label{fig:nanowire_current}
\end{figure}

In this section, we analyze the frequency tuning and Hamiltonian of the system when exposed to an external magnetic field. As shown in Fig. \ref{fig:nanowire_current}, with an external magnetic field $B_{\textrm{ext}}$, the DC supercurrent is expressed as
\begin{equation}
    I_{\textrm{dc}} = \frac{B_{\textrm{ext}} d}{L_{\textrm{dc}}},
\end{equation}
where \( d \) is the gap between the two nanowires, and \( L_{\textrm{dc}} \) is the effective geometric inductance per unit length. The kinetic inductance of the two nanowires are modified as follows
\begin{equation}
\begin{aligned}
     L_{k,1}(I) &= L_{1} \left[1 + \frac{(I_{\mathrm{dc}} + I_{\mathrm{rf1}})^2}{I_{1}^{*2}}\right], \\
     L_{k,2}(I) &= L_{2} \left[1 + \frac{(I_{\mathrm{dc}} - I_{\mathrm{rf2}})^2}{I_{2}^{*2}}\right].
\end{aligned}
\end{equation}
where \( L_{1(2)} \) denotes the kinetic inductance of each nanowire, \( I_{\mathrm{rf1}} \) is the RF current, and \( I_{1(2)}^* \) represents a characteristic current on the order of the critical current. These terms are related via the width ratio \( \gamma = w_{2}/w_{1} \) as
\begin{equation}
\begin{aligned}
    L_{2} &= \frac{L_{1}}{\gamma}, \\
    I_{\mathrm{rf2}} L_{k,2}(I_\mathrm{dc}) &= I_{\mathrm{rf1}}L_{k,1}(I_\mathrm{dc}), \\
    I_{2}^* &= \gamma I_{1}^*.
\label{equ:relations}
\end{aligned}
\end{equation}
The total inductance is given by:
\begin{equation}
\begin{aligned}
    \frac{1}{L_{k}(I)} &= \frac{1}{L_{k,1}(I)} + \frac{1}{L_{k,2}(I)} \\
        &= \frac{1}{\gamma L_{2}} \left[1 + \left(\frac{\gamma I_{\textrm{dc}} + I_{\textrm{rf2}}}{I_{2}^*}\right)^2\right]^{-1} \\
        & + \frac{1}{L_{2}} \left[1 + \left(\frac{ I_{\textrm{dc}} - I_{\textrm{rf2}}}{I_{2}^*}\right)^2\right]^{-1}.
\end{aligned}
\end{equation}
Without considering the RF current, the frequency shift can thus be expressed as
\begin{equation}
    \frac{\Delta f}{f} = \frac{}{} = -\frac{\gamma}{2} \frac{I_{\textrm{dc}}^2}{I_{2}^{*2}}.
\end{equation}
Let \( w_{2} < w_{1} \); then the maximum frequency shift corresponds to the maximum DC supercurrent that can be applied to the thinner nanowire before superconductivity breaks down. According to \cite{xu2019frequency}, the maximum frequency shift is around 3\%. In our case, with \( \gamma = 0.5 \), the maximum frequency shift is approximately 1.5\%. 

The potential energy stored in the inductance can be written as
\begin{equation}
E_L(I) = \frac{1}{2} L_{k, 1} (I_{\mathrm{rf1}} + I_{\mathrm{dc}})^2 + \frac{1}{2} L_{k, 2} (I_{\mathrm{rf2}} - I_{\mathrm{dc}})^2.
\end{equation}
One can use Eq. \ref{equ:relations} to expand the above equation on the orders of $I_\mathrm{rf2}$
\begin{equation}
E_L(I) = \dots + T \cdot L_2 I_{\mathrm{rf2}}^3 + F \cdot L_2 I_{\mathrm{rf2}}^4,
\end{equation}
where the coefficients \(T\) and \(F\), representing the three-wave mixing (TWM) and four-wave mixing (FWM) effects, respectively, are given by
\begin{equation}
\begin{aligned}
    T & = \frac{2 I_{\text{dc}} L_2}{I^{*2}} \left( \frac{\left( I_{\text{dc}}^2 + I^{*2} \right)^3}{\left( \gamma^2 I_{\text{dc}}^2 + I^{*2} \right)^3} - 1 \right), \\
    F & = \frac{L_2}{2I^{*2}} \left( 1 + \frac{\left( I_{\text{dc}}^2 + I^{*2} \right)^4}{\gamma \left( \gamma^2 I_{\text{dc}}^2 + I^{*2} \right)^4} \right).
\end{aligned}
\end{equation}
The three-wave mixing components are non-zero when $\gamma \neq 1$, indicating the asymmetric widths of the two nanowires. Therefore, we intentionally design the widths to be asymmetrical.

\begin{figure}[t]
    \centering
    \includegraphics[width=\linewidth]{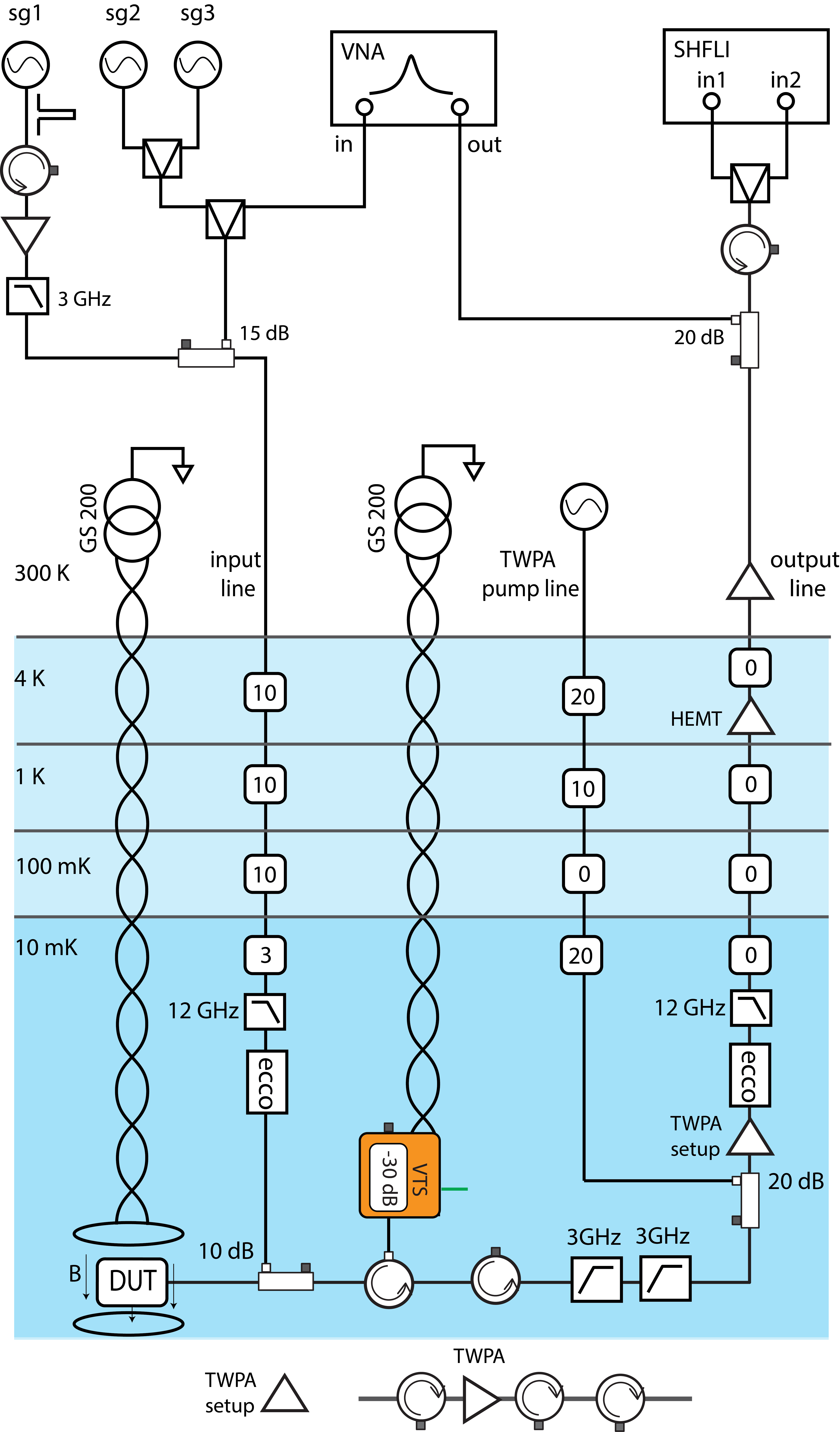}
    \caption{Illustration of the measurement setup and the cryogenic wiring for frequency conversion and noise calibration. The pump and probe tones are combined before being sent to the input line. They are then coupled to the device under test (DUT) via a 10 dB directional coupler. The DUT is positioned inside a homemade coil for applying magnetic field. The signal tone, generated idler tone, and pump tone are reflected to the output line, where the pump tone is filtered by low-pass filters. The output microwaves are pre-amplified by a TWPA and then further amplified by a HEMT amplifier. A variable temperature stage (VTS) is used to source thermal noise to the DUT for noise calibration.}
    \label{fig:app_measSetup}
\end{figure}

\section{Measurement Setup and Noise Calibration}
\label{sec:app_noiseCalibration}
The measurement setup is illustrated in Fig. \ref{fig:app_measSetup}. The pump signal, ``sg1," is generated by Zurich Instrument SHFSG, which produces the pulsed waveform. The probe tones, ``sg2" and ``sg3," are generated by Zurich Instrument SHFLI. These probe tones are combined with the probe tone of a Vector Network Analyzer (VNA) and then sent to the input line. The combined tones are attenuated and coupled to the resonator using a 10 dB directional coupler. The reflected microwaves, including the signal tone, generated idler tone, and pump tone, are directed to the output line, where the pump tone is filtered out by low-pass filters.

The output microwaves are initially amplified by a Traveling-Wave Parametric Amplifier (TWPA) and subsequently by a High-Electron-Mobility Transistor (HEMT) amplifier. The output is split by a 20 dB directional coupler, with the microwave from the coupling port sent to the VNA, and the through port tone sent to the SHFLI for measurement. Both SHFSG and SHFLI are limited to a maximum frequency output of 8.5 GHz, which restricts the measurement bandwidth of our setup.

To calibrate the added noise of the conversion process, we utilize a Voltage-Temperature Source (VTS) to generate thermal noise directed at the Device Under Test (DUT). This thermal noise is then reflected back to the output line. To eliminate the added noise introduced by the pre-amplifier, we measure the total noise added by the circuit, referenced to the VTS, with the conversion tone both on and off. The added noise from the conversion process is determined by calculating the difference between these two measurements, after accounting for the loss from the VTS to the TWPA.
\begin{figure}[t]
    \centering
    \includegraphics[width=0.9\linewidth]{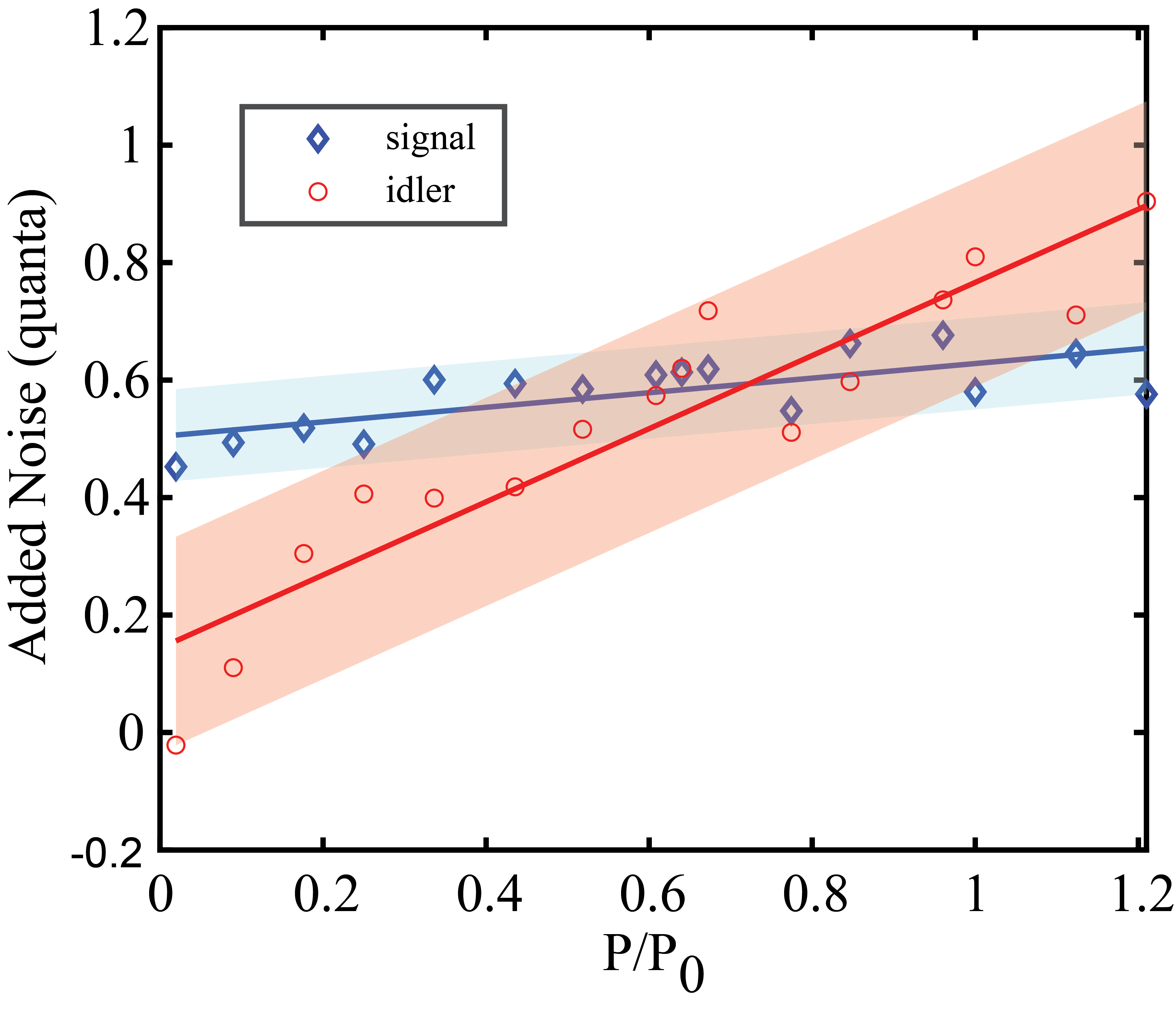}
    \caption{Added noise of the signal mode (blue diamond) and the idler mode (red circle) with respect to the normalized pump power. The solid line represent the corresponding linear fitting.}
    \label{fig:addedNoise}
\end{figure}

The added noise, calibrated and presented in Fig.~\ref{fig:addedNoise}, shows a linear trend with pump power for both the signal and idler. The slopes are fitted as 0.04 and 0.07, respectively. The discrepancy between the two can be attributed to energy transfer due to conversion and possible coupling to other modes. At near-unity conversion effciency, where the normalized pump power $P/P_0 = 1$, the added noise for the signal and the idler modes are $0.59\pm0.04$ and $0.79\pm0.08$ quanta, respectively.

\section{Conversion at the Single-photon Level}
\label{sec:app_SinglePhotonConversion}
\begin{figure}[t]
    \centering
    \includegraphics[width=0.85\linewidth]{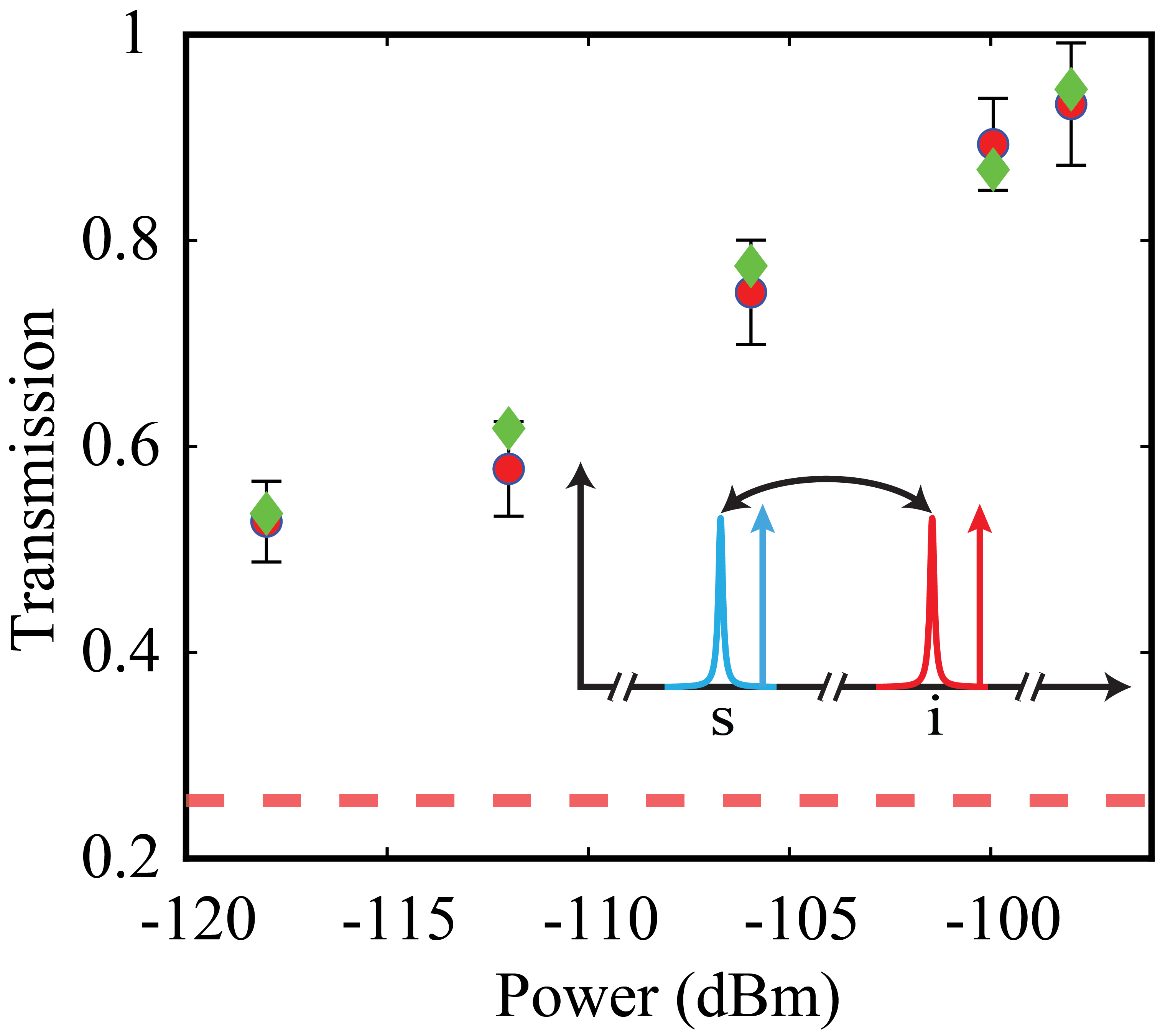}
    \caption{The conversion efficiency as a function of the power of the TLS saturation tone. The inset illustrates the process: the saturation tones are detuned from the corresponding signal or idler tone by 2 MHz, and are sufficiently filtered before the measurement setup. The red dashed line represents the reference efficiency without the saturation tone applied. The green diamonds and the red circles are the total coupling efficienct and the measured conversion efficiency, respectively.}
    \label{fig:app_singlePhotonConversio}
\end{figure}

Two-level systems (TLSs) are a major source of loss in superconducting quantum circuits \cite{pappas2011two, muller2019towards}. These TLSs likely originate from the oxide layer formed during the fabrication process \cite{crowley2023disentangling}. At high probe power, the TLSs become saturated, resulting in a higher internal quality factor. Conversely, at low probe power, the resonator couples with the TLSs, leading to a degraded internal quality factor. One can saturate the two-level system with a pump tone that is detuned from the target mode \cite{andersson2021acoustic}. With appropriate detuning ($> 3$ linewidth), we could improve the quality factor of the resonator while the pump tone could be sufficiently filtered at the measurement setup. 

Fig. \ref{fig:app_singlePhotonConversion} illustrates the conversion efficiency with (red circles) and without (red dashed line) the TLS saturation tone. The green diamonds represents the total coupling efficiency. At the probe power of -148 dBm, which corresponding to a single intra-cavity photon, the conversion efficiency is 26\%. With the applied TLS saturation tone, the increased internal quality factor result in a increasing total coupling efficiency and therefore a higher on-chip conversion efficiency. At the TLS saturation pump power of -98 dBm, we achieve a 93\% conversion efficiency. 

It is worth noting that the degree of TLS-induced loss varies across different devices and fabrication processes. While our measurements show that a TLS saturation pump can substantially improve conversion efficiency, the actual gains may differ depending on factors such as film quality, substrate preparation, and packaging. To effectively apply this method, the resonance linewidth must be significantly smaller than the TLS linewidth (on the order of MHz \cite{andersson2021acoustic, kristen2023observation}). Moreover, the TLS saturation pump should be adequately detuned to minimize backaction on the conversion process and carefully filtered to prevent interference with the measured signals.

\begin{figure*}[t]
    \centering
    \includegraphics[width=\textwidth]{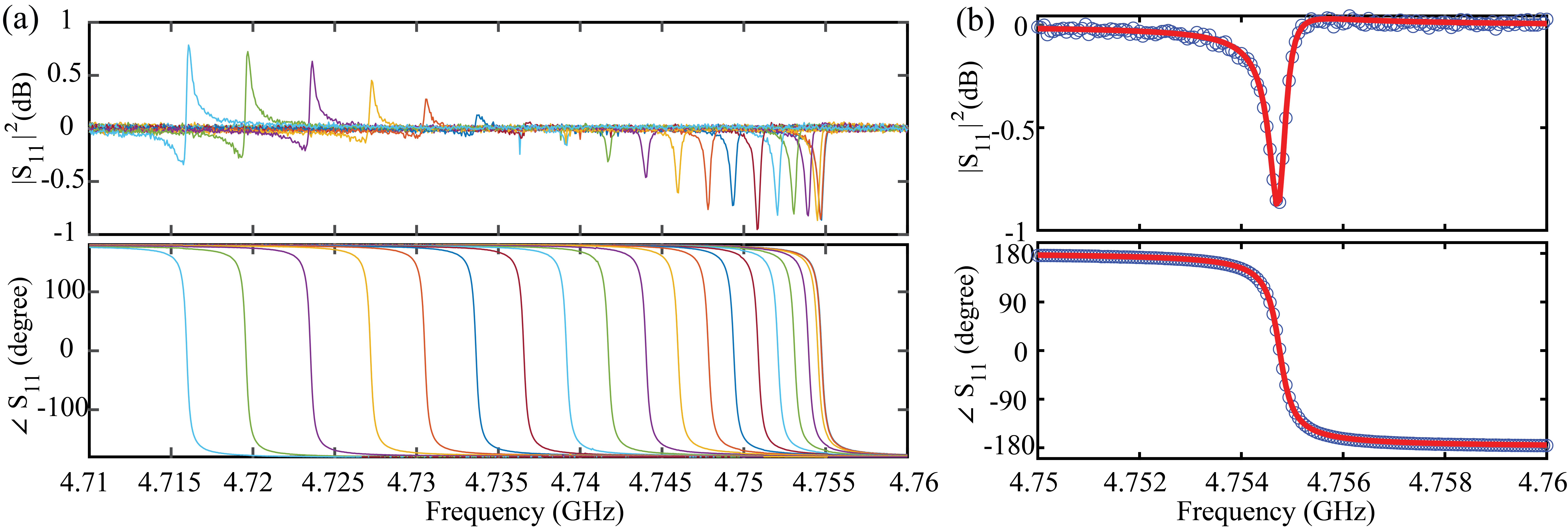}
    \caption{(a) Reflection spectrum of the signal mode under tuning magnetic field. Top panel and the bottom panel are the magnitude and phase of the reflection spectrum, respectively. (b) The quality factor fitting of the resonator. }
    \label{fig:app-frequencyShift}
\end{figure*}

\section{Qualify Factor Under Tuning Magnetic Field}
\label{sec:app_qWithMagFieldChange}
In this section, we provide a detailed analysis of the quality factor changes depicted in Fig. \ref{fig:2-frequency_tuning}(c) of the main text. We focus on the signal mode at 4.75 GHz used in the conversion process, and monitor how its resonance varies with the tuning of the magnetic field. The resonance shift under the tuning magnetic field is shown in Fig. \ref{fig:app-frequencyShift} (a). With a higher magnetic field, the resonance frequency shift to the left, while the resonance dip become shallower and eventually diminishes at around 4.74 GHz. With a even higher magnetic field, the resonance starts to have a Fano shape, with asymmetric resonance profile, which is likely due to coupling with the package modes.

This phenomenon appears to contradict the findings in Refs.~\cite{samkharadze2016high, kroll2019magnetic, yu2021magnetic}, where quality factor is shown to decrease with increasing magnetic field. However, the magnetic field applied in our study is sufficiently small (\(< 0.2\) mT, compared to several hundred mT in these references) that it does not induce any noticeable degradation in \( Q \). The wireless current injection approach also minimize the radiation loss induced by the dc coupling line as shown in Ref. \cite{mahashabde2020fast}.

The increase in the quality factor can be attributed to reduced coupling to two-level systems (TLSs) caused by the frequency shift. TLSs originating from surface oxides and interfaces can be both spatially and spectrally localized \cite{barends2013coherent, lisenfeld2019electric}. The shift in the resonant frequency due to the applied magnetic field can modify the coupling by affecting both the mode profile and the operating frequency. 

\section{Amplification Characterization}
\label{sec:app_amplification}
In this appendix, we provide additional details on the amplification performance of our multimode device under different operating conditions. These measurements confirm that our high-kinetic-inductance NbN resonator can function as both a four-wave mixing and three-wave mixing amplifier.

\begin{figure*}[t]
    \centering
    \includegraphics[width=\textwidth]{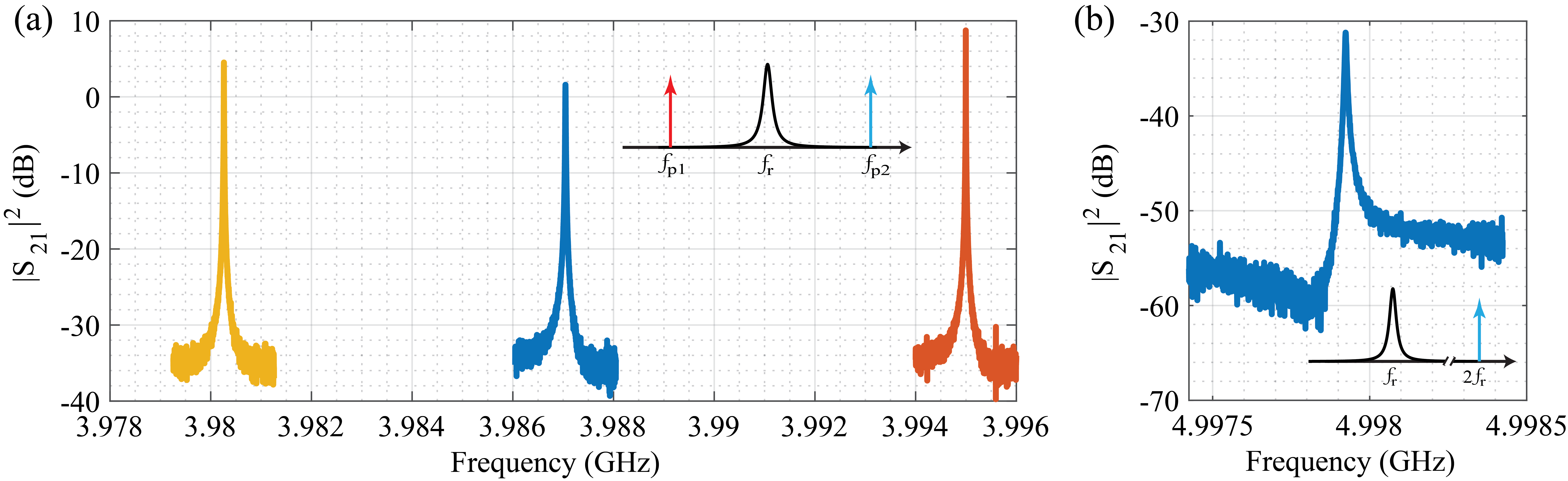}
    \caption{Characterization of Amplification Performance. (a) Four-wave mixing amplification at 3 K under different magnetic fields (from right to left: 0 mT, 0.09 mT, and 0.13 mT). (b) Three-wave mixing amplification at 10 mK. The insets illustrate the corresponding pump schemes.}
    \label{fig:app_gain}
\end{figure*}

During preliminary screening at 3 K, we focused on the resonance near 3.99~GHz and tuned it using an external magnetic field. Under three distinct magnetic field settings, we employed a two-tone pump scheme (Fig.~\ref{fig:app_gain}~(a) inset), with each pump tone detuned from the resonance center by more than three linewidths. This configuration yielded gains exceeding 35~dB across the tuned frequency range as shown in Fig.~\ref{fig:app_gain}~(a).

We also demonstrated amplification via a three-wave mixing Hamiltonian. At 10 mK, We apply magnetic field (0.15 mT) to activate the three-wave mixing nonlinearity of the device. For the mode centered at 4.998~GHz, we applied a pump at twice the resonance frequency (9.995~GHz), achieving a gain of more than 20~dB, as shown in Fig.~\ref{fig:app_gain}~(b).

These observations confirm that our device supports robust amplification under varying conditions, underscoring its potential versatility for applications in quantum information processing. We anticipate that further optimization of device parameters, pump configuration, and operating temperatures will enable even higher gains and broader bandwidths in future iterations. 

\bibliography{apssamp}

\end{document}